\documentclass[twocolumn,showpacs,preprintnumbers,amsmath,amssymb]{revtex4-1}

\usepackage{graphicx}
\usepackage{bm}
\usepackage{amsmath}

\begin{document}

\title{Concentration fluctuations and phase transitions in coupled
modulated bilayers}

\author{Yuichi Hirose}
\author{Shigeyuki Komura}
\email{komura@tmu.ac.jp}
\affiliation{%
Department of Chemistry,
Graduate School of Science and Engineering,
Tokyo Metropolitan University, Tokyo 192-0397, Japan
}%
\author{David Andelman}
\affiliation{%
Raymond and Beverly Sackler School of Physics and Astronomy,
Tel Aviv University, Ramat Aviv, Tel Aviv 69978, Israel
}%


\begin{abstract}
We consider the formation of finite-size domains in
lipid bilayers consisting of saturated and hybrid lipids.
First, we describe a monolayer model that includes a coupling
between a compositional scalar field and a two-dimensional vectorial
order-parameter.
Such a coupling yields an effective two-dimensional microemulsion
free-energy for the lipid monolayer, and its characteristic length of
compositional modulations can be considered as the origin of finite-size domains in
biological membranes.
Next, we consider a coupled bilayer composed of two modulated monolayers,
and discuss the static and dynamic properties of concentration
fluctuations above the transition temperature.
We also investigate the micro-phase separation below the transition
temperature, and compare the micro-phase separated structures with
statics and dynamics of concentration fluctuations above the transition.
\end{abstract}

\maketitle

\section{Introduction}
\label{sec:introduction}

Biomembranes are two-dimensional (2D) fluids that separate the inner
and outer environment of organelles in biological cells.
Naturally occurring
biomembranes  consist typically of numerous lipid
species, sterols, sugars and membrane proteins.
According to the ``lipid raft'' hypothesis~\cite{SI}, some of the lipid components
and/or proteins are incorporated into finite-size domains, which play
an important role on cellular functions such as signal transduction
processes.
Recent experiments suggest that lipid rafts are nothing but dynamical
molecular assemblies of 20\,nm in size with finite lifetime
in the order of 10--20\,ms~\cite{ERMSSPBHMSH}.

Being motivated by the raft hypothesis, a large number of investigations~\cite{DBVTLJE,VK,VK3,VK2,BHW,RKG,KG,KGAHWF,VSKG,SCCVNSK,SVK,VCSSHB}
have been conducted to reveal the properties of artificial
membranes consisting of lipid mixtures and cholesterol.
Below the miscibility transition temperature, formation of micron-size
domains were observed using fluorescent
microscopy~\cite{DBVTLJE,VK,VK3,VK2}.
In some cases, rather than a macroscopic phase separation, domains with
distinct size in the micrometer range have been reported~\cite{BHW,RKG,KG}.
For example, various types of modulated (stripe or hexagonal) patterns have
been found for multicomponent lipid and cholesterol mixtures~\cite{KGAHWF}.
Above the miscibility transition temperature,
even multicomponent membranes do not phase separate, and their concentration
fluctuations around the homogeneous state can be investigated~\cite{VSKG,SCCVNSK,SVK},
in particular, close to the critical point, $T_{\rm c}$.
Furthermore, it is interesting to note that critical concentration fluctuations have been
observed in membranes extracted from living cells~\cite{VCSSHB}.

One of the main reasons that initiated the notion of lipid rafts in biomembranes
is the existence of finite-size domains, rather than domains resulting from a
macroscopic phase separation~\cite{HaatajaFEBS}.
Assuming that membranes are in equilibrium, the same question can be phrased
in terms of whether the biomembrane state is above or below the
miscibility transition temperature.
In the high temperature one-phase region, the only relevant length
scale is the correlation length associated with concentration fluctuations,
and it diverges at the critical point, $T_{\rm c}$~\cite{footnote1}.
Below this temperature, there should be a physical mechanism
suppressing domain coarsening in order to explain the existence
of finite-size domains
in equilibrium.

Yet another characteristic feature of biomembranes is that the
lipid composition of the two leaflets (monolayers) constituting the bilayer
is not the same~\cite{AZA}.
Moreover, such asymmetric monolayers are not independent
but are coupled to one another.
This was  confirmed experimentally by investigating the phase separation
of bilayers with different monolayer lipid compositions~\cite{CK},
or seen in simulations~\cite{Perlmutter}.
One mechanism that leads to the coupling between the two leaflets is the lipid chain
interdigitation occurring at the mid-plane of the bilayer~\cite{WLM,PS,May},
which may affect the domain size in asymmetric bilayers.

One possibility to account for such finite-size domains in lipid mixtures
is to consider the special role played by
``hybrid lipids''~\cite{BPS,BS,YBS,YS}.
These lipids have one saturated hydrocarbon chain and another unsaturated one.
They tend to be localized at boundaries of 2D domains and act as a line-active agent.
Hence, a ternary lipid mixture consisting of saturated lipids with two
saturated chains, unsaturated lipids with two unsaturated chains,
and hybrid lipids can be regarded as a {\it ``2D microemulsion"}, because it
is analogous to the microemulsion phase consisting of
oil, water and surfactant~\cite{SV,SVK}.
For microemulsions it is known that there is another length scale,
in addition to the correlation length mentioned above,
related to the size of water and oil micro-domains~\cite{GSbook}.

Using the microemulsion analogy, we suggest that an additional
length scale can be responsible for the finite-size structures in
biomembranes. We propose a model for bilayers consisting
of two coupled monolayers that are both in a 2D microemulsion state,
manifesting modulated phases.
First, we describe a monolayer model that includes a coupling between
the lipid composition and a 2D vectorial order-parameter. The model accounts for the
line active nature of hybrid lipids.
Next, we consider a bilayer in which two such monolayers
are coupled through an inter-leaflet interaction.
For the coupled bilayer, we discuss
the static and dynamic properties of concentration fluctuations
above $T_{\rm c}$,
and investigate their micro-phase separation below $T_{\rm c}$~\cite{HKA}.
Intermediate structures arise when two competing structures have
different characteristic length scales.
One of our important conclusions is that the micro-phase separated structures below
$T_{\rm c}$ reflect the static and dynamic properties of concentration
fluctuations above $T_{\rm c}$.

This paper is organized as follows.
In Sec.~\ref{sec:model}, after describing how the microemulsion state
is obtained for monolayers, a model for coupled modulated bilayers is
presented.
In Sec.~\ref{sec:fluctuation}, we show the results of static
and dynamic structure factors for the coupled bilayers above
$T_{\rm c}$.
In Sec.~\ref{sec:separation}, we describe some results for competing
micro-phase separation in coupled bilayers below $T_{\rm c}$.
Finally, in Sec.~\ref{sec:discussion}, further discussion
and final remarks are presented.

\section{Model}
\label{sec:model}

\subsection{Modulated lipid monolayers}

\begin{figure}[tbh]
\begin{center}
\includegraphics[scale=0.25]{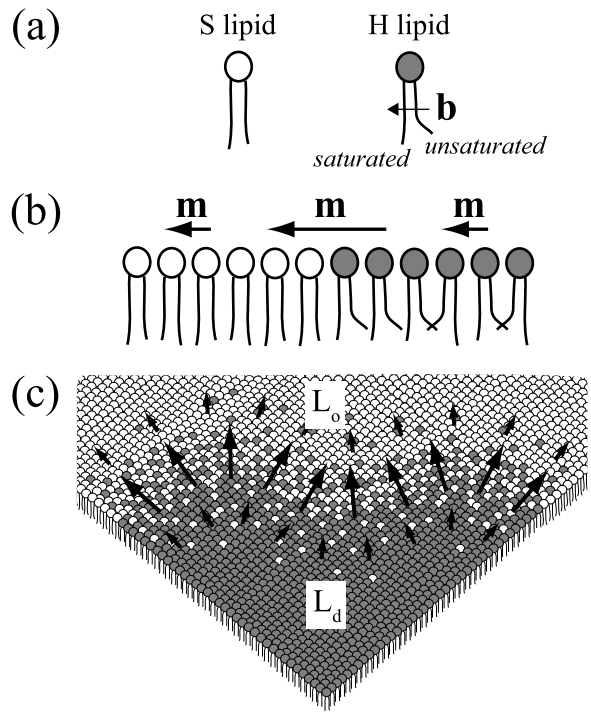}
\end{center}
\caption{
(a) Saturated lipids (S) with two saturated hydrocarbon chains, and
hybrid lipids (H) with one saturated and another unsaturated chain.
For each hybrid lipid, we define a 2D vector ${\mathbf b}$ pointing
from its unsaturated tail toward the saturated one.
(b) The vector $\mathbf{m}$ denotes a coarse-grained 2D
vectorial order-parameter over the microscopic $\mathbf{b}$ vectors.
(c) A binary lipid monolayer consisting of saturated and hybrid lipids.
At the interface between the ${\rm L_o}$ and ${\rm L_d}$ phases,
hybrid lipids orient their saturated chains toward the ${\rm L_o}$ phase.
The distribution of the $\mathbf{m}({\mathbf r})$ vector field is shown
by arrows of variable length.
}
\label{fig1}
\end{figure}

We consider a 2D microemulsion formation in a monolayer consisting
of two types of lipids: a saturated lipid (denoted by ``S'') and
a hybrid one (``H'').
As shown schematically in Fig.~\ref{fig1}(a), the saturated lipid has two saturated
chains, whereas the hybrid lipid  has one saturated chain
and an unsaturated one. It is well known~\cite{DBVTLJE,VK,VK3,VK2} that the resulting
liquid-ordered (${\rm L_o}$) and liquid-disordered (${\rm L_d}$)
phases are rich in  saturated and  hybrid lipids, respectively, as is
depicted in Fig.~\ref{fig1}(c).
In the experiments mentioned previously, cholesterol is usually added as a
third component, and is known to affect the area per headgroup of
lipids~\cite{EN,HLCH}.
However, since cholesterol has a strong preference for the
${\rm L_o}$ phase and affects mainly the saturated lipid, we do not
consider cholesterol explicitly in our model, and neglect its presence hereafter.

The local area fraction of saturated and hybrid lipids are defined as
$\phi_{\rm S}(\mathbf{r})$ and $\phi_{\rm H}(\mathbf{r})$,
respectively, where $\mathbf{r}=(x,y)$ is a 2D vector.
Under the incompressibility condition,
$\phi_{\rm S}+\phi_{\rm H}=1$, and
the only relevant order-parameter is the difference between the two lipid
compositions, $\phi=\phi_{\rm S}-\phi_{\rm H}$.
The phenomenological free-energy that describes the 2D phase-separation
between S and H lipids is given by a Ginzburg-Landau expansion in terms of
the order-parameter $\phi$:
\begin{equation}
F_{\rm s}[\phi] = \int {\rm d}^2r \,
\left[ \frac{\sigma}{2}(\nabla \phi)^2 + \frac{\tau}{2} \phi^2
+ \frac{1}{4} \phi^4 - \mu \phi \right].
\label{freeenergygl}
\end{equation}
Here $\sigma >0$ is related to the line tension between monolayer
domains, $\tau \propto T-T_{\rm c}$ is the reduced temperature with
respect to the critical point $T_{\rm c}$, and $\mu$ is the chemical
potential that regulates the average $\phi$ value in the monolayer.
Without loss of generality, the coefficient of the quartic term is
set to be a positive constant by an appropriate rescaling of the position variable,
$\mathbf{r}$.

Next we discuss the role of the hybrid lipid and its effect on the phase
separation.
We define a 2D lateral vector ${\mathbf b}$ pointing from the unsaturated
tail  of the hybrid lipid toward its saturated one, as depicted in
Fig.~\ref{fig1}(a).
We then introduce a coarse-grained 2D vectorial order-parameter field
$\mathbf{m}({\mathbf r})=(m_x({\mathbf r}),m_y({\mathbf r}))$, as shown
in Fig.~\ref{fig1}(b).
This vector is the spatial average of the ${\mathbf b}$ vectors over areas large
as compared with molecular size, but still small enough as compared with
macroscopic scales.
The phenomenological free-energy associated with the vectorial field $\mathbf{m}$
(and due to the
hybrid lipids) can be written as an expansion up to quadratic order in
$\mathbf{m}$~\cite{CLM}:
\begin{equation}
F_{\rm h}[\mathbf{m}] = \int {\rm d}^2 r \,
\left[ \frac{K}{2}(\nabla \cdot \mathbf{m})^2 +
\frac{a}{2} \mathbf{m}^2 \right].
\label{freeenergyhybrid}
\end{equation}
The coefficient $K$ is the 2D elastic constant, while $a$ is taken
to be positive so that $\mathbf{m}=0$  is the stable homogeneous state.
An additional term  $(\nabla \times \mathbf{m})^2$ is allowed by symmetry,
but is not considered at present for simplicity sake.

At the ${\rm L_o}/{\rm L_d}$ interface, hybrid lipids orient their
saturated and unsaturated chains toward the ${\rm L_o}$  and
${\rm L_d}$ phases, respectively, thereby reducing the chain
mismatch~\cite{YBS,YS} as shown in Fig.~\ref{fig1}(c).
In fact, the line activity of hybrid lipid was predicted~\cite{YBS,YS}
to be more pronounced for binary mixtures of saturated and hybrid lipids
as compared to ternary mixtures.
Within our phenomenological approach, the role of hybrid lipids can
be represented by a coupling between the lateral variation of
$\phi$ and the vectorial field $\mathbf{m}$.
To lowest orders, the coupling term in the free energy is given by
\begin{equation}
F_{\rm c}[\phi, \mathbf{m}] =
- \Gamma \int {\rm d}^2 r \, \mathbf{m} \cdot (\nabla \phi),
\label{freeenergycoup}
\end{equation}
where $\Gamma$ is a positive coupling constant
because the vector $\mathbf{m}$ tends to orient towards the
${\rm L_o}$ domains.
The total monolayer free-energy is given here by the sum of the
three terms introduced in Eqs.~(\ref{freeenergygl})--(\ref{freeenergycoup}):
$F_{\rm m} = F_{\rm s}+ F_{\rm h}+ F_{\rm c}$.
Notice that our model is valid in the weak segregation limit (close
to $T_{\rm c}$), because slow spatial variation
of the order parameters is intrinsically assumed.
We further note that a similar free energy functional using a vectorial order-parameter was
proposed for 3D microemulsions~\cite{CJPW}.

The total free energy can be conveniently expressed using the 2D Fourier
transform of $\phi$
\begin{equation}
\phi(\mathbf{q})= \int {\rm d}^2 r\,
\phi(\mathbf{r}) e^{-i \mathbf{q} \cdot \mathbf{r}},
\end{equation}
where $\mathbf{q}=(q_x,q_y)$ is a 2D wavevector, and
similarly $\mathbf{m}(\mathbf{q})$ is the Fourier transform of
$\mathbf{m}(\mathbf{r})$.
Minimizing $F_{\rm m}$ with respect to ${\mathbf m}$, we obtain its
optimum value as
\begin{equation}
\mathbf{m}(\mathbf{q}) = \frac{i \Gamma \mathbf{q}}{a+K q^2}
\phi(\mathbf{q}),
\label{eqm}
\end{equation}
where $q = \vert \mathbf{q} \vert$.
By substituting back Eq.~(\ref{eqm}) into the monolayer free energy
$F_{\rm m}$, the minimized free energy is expressed as
\begin{align}
F_{\rm m}[\phi]  = &
\int \frac{{\rm d}^2 q}{(2\pi)^2} \,
\frac{1}{2} \left[ \tau+\sigma q^2
-\frac{\Gamma^2 q^2}{a+K q^2} \right] \phi(\mathbf{q}) \phi(-\mathbf{q})
\nonumber \\
& + \int {\rm d}^2 r \,
\left[ \frac{1}{4} \phi^4 - \mu \phi \right],
\label{minfourierenergy}
\end{align}
where the Fourier transform has been used only for the second-order
terms in $\phi$.
Expanding the effective binary interaction for small $q$,
we obtain
\begin{align}
F_{\rm m}[\phi] \approx &
\int \frac{{\rm d}^2 q}{(2\pi)^2} \,
\left[
2 B q^4 - 2 Aq^2 + \frac{\tau}{2} \right] \phi(\mathbf{q}) \phi(-\mathbf{q})
\nonumber \\
& + \int {\rm d}^2 r \,
\left[ \frac{1}{4} \phi^4 - \mu \phi \right],
\label{fphim3}
\end{align}
where two new parameters are defined
\begin{equation}
\label{e8}
B \equiv \frac{K \Gamma^2}{4a^2},~~~~~
A \equiv \frac{1}{4} \left( \frac{\Gamma^2}{a} - \sigma \right).
\end{equation}
When the coupling constant $\Gamma$ is small enough, i.e.,
$\Gamma^2/a<\sigma$ (or $A<0$), the minimum of the Gaussian term
(first integral) in Eq.~(\ref{fphim3}) occurs at $q^{\ast}=0$,
and is a signature of a macroscopic phase separation (for $\tau<0$).
On the other hand, when $\Gamma$ is large enough, i.e., $\Gamma^2/a>\sigma$ (or
$A>0$), the minimum occurs at a non-zero wavenumber $q^{\ast} \neq 0$
given by
\begin{equation}
q^{\ast} = \sqrt{\frac{A}{2B}}
= \sqrt{ \frac{a(1-a \sigma /\Gamma^2)}{2 K}},
\label{qcharac}
\end{equation}
indicating a potential micro-phase separation with $q^{\ast}>0$ modulation.

Taking the inverse Fourier transform, Eq.~(\ref{fphim3}) can be expressed in
position space as
\begin{align}
F_{\rm m}[\phi] = \int {\rm d}^2 r \,
& \biggl[ 2 B ( \nabla^2 \phi )^2 - 2 A ( \nabla \phi )^2
\nonumber \\
& + \frac{\tau}{2} \phi^2 + \frac{1}{4}\phi^4 - \mu \phi \biggr].
\label{mono_minimized}
\end{align}
This is the so-called ``2D microemulsion" free-energy for a monolayer composed
of a binary lipid mixture.
When $A>0$, the negative gradient-squared term favors spatial modulations,
while the positive Laplacian squared term with $B>0$
suppresses modulations.
As mentioned above, Yamamoto {\it et al.}~\cite{YBS,YS} showed that the effective
line tension between domains becomes negative for membranes consisting
of saturated and hybrid lipids.

More generally, models based on equations similar to Eq.~(\ref{mono_minimized})
have been used successfully in the past to describe modulated phases~\cite{SA} arising in a
variety of different biophysical and chemical systems such as Langmuir
films~\cite{ABJ}, lipid membranes~\cite{LA,KSA} and diblock
copolymers~\cite{Leibler}.

\subsection{Coupled lipid bilayers}
\label{couplebilayer}

\begin{figure}[tbh]
\begin{center}
\includegraphics[scale=0.18]{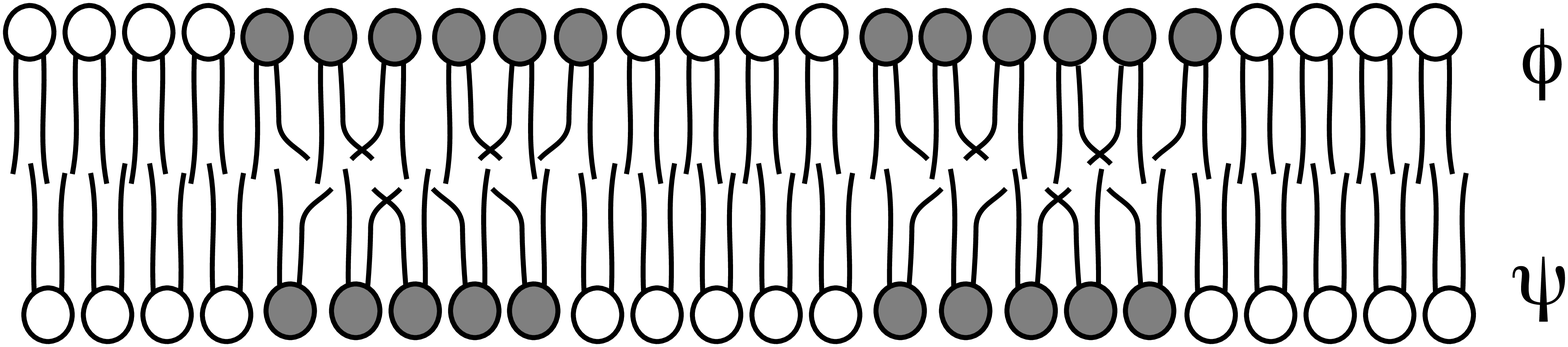}
\end{center}
\caption{
Schematic illustration of two coupled modulated monolayers forming
a bilayer membrane.
Each monolayer is composed of a binary S/H lipid mixture, which can have
a lateral composition modulation.
The relative composition of S and H lipids in the upper and lower
leaflets are defined by $\phi$ and $\psi$, respectively.
The lipid tails interact across the bilayer mid-plane.
}
\label{fig1.5}
\end{figure}

We consider two coupled modulated monolayers forming a bilayer as shown
in Fig.~\ref{fig1.5}.
Each monolayer is a binary mixture of saturated and hybrid lipids.
We define two local
order-parameters for the two monolayers:
$\phi=\phi_{\rm S}-\phi_{\rm H}$ and
$\psi=\psi_{\rm S}-\psi_{\rm H}$, depicted in Fig.~\ref{fig1.5}.
The coarse-grained free-energy functional is then written as~\cite{HKA}:
\begin{align}
F_{\rm b}[\phi, \psi ] = \int {\rm d}^2r
& \biggl[ 2 B( \nabla^2 \phi)^2 - 2 A(\nabla \phi)^2
\nonumber \\
& + \frac{\tau_{\phi}}{2}\phi^2 + \frac{1}{4}\phi^4 - \mu_{\phi}\phi
\nonumber \\
& + 2 D( \nabla^2 \psi)^2 - 2 C( \nabla \psi)^2
\nonumber \\
& + \frac{\tau_{\psi}}{2}\psi^2 + \frac{1}{4}\psi^4 - \mu_{\psi}\psi
- \Lambda \phi \psi \biggr].
\label{freeenergy}
\end{align}
The first five terms depend only on $\phi$ and its derivatives and
describe the upper monolayer in Fig.~\ref{fig1.5} and its possible
modulations. These are the same terms as in Eq.~(\ref{mono_minimized}) for the single monolayer case.
Similarly, the latter five $\psi$ terms describe the lower monolayer, where $\tau_{\phi}$ and $\tau_{\psi}$ are the two reduced
temperatures, while $\mu_{\phi}$ and $\mu_{\psi}$ are the corresponding chemical
potentials.
The last term, $\Lambda \phi \psi$, represents the coupling
between the two leaflets with a coupling constant $\Lambda$,

We would like to comment on the physical origin of the proposed coupling
term, $-\Lambda \phi \psi$, in Eq.~(\ref{freeenergy}).
Note that this quadratic term is invariant under exchange
of $\phi \leftrightarrow \psi$.
When $\Lambda>0$, this term can be obtained from a term such as
$(\phi-\psi)^2$, which represents a local energy penalty when the
upper and lower monolayers have different compositions~\cite{WLM,PS}.
For mixed lipid bilayers, such a coupling may result from
the conformational confinement of the lipid chains, and hence would
have an entropic origin~\cite{WLM}.
By estimating the degree of lipid chain interdigitation, the
order of magnitude of the coupling parameter $\Lambda$ can be evaluated~\cite{May}.
In general, $\Lambda$ can also be negative
depending on the specific coupling mechanism~\cite{SKA}.
However, as the sign of $\Lambda$ does not affect our result, it
is sufficient to consider the $\Lambda>0$ case.
The other possible higher-order coupling terms which are allowed by
symmetry are $\phi^2 \psi^2$, $\phi \psi^3$ and $\phi^3 \psi$.
However these terms do not affect the properties of concentration
fluctuations in any essential way, and will not be considered hereafter.

\section{Concentration fluctuations above $T_{\rm c}$}
\label{sec:fluctuation}

Using the bilayer free-energy, Eq.~(\ref{freeenergy}), we obtain the static
and dynamic structure factors, which describe the properties of concentration
fluctuations for two coupled monolayers in
their respective one-phase region (above $T_{\rm c}$).
We shall closely follow the formulation of Ref.~\cite{OBCK} in which
the coupled macro-phase separation was discussed for bilayers.

\subsection{Static structure factor}

The spatially varying $\phi({\mathbf r})$ and $\psi({\mathbf r})$ can
be written as
$\phi({\mathbf r}) = \phi_{\rm 0} + \delta \phi({\mathbf r})$ and
$\psi({\mathbf r}) = \psi_{\rm 0} + \delta \psi({\mathbf r})$,
respectively, where $\phi_0=\langle \phi \rangle$ and $\psi_0=\langle \psi \rangle$
are the spatially averaged monolayer compositions, and
$\delta \phi$ and $\delta \psi$ describe the deviations from their
average values.
In thermal equilibrium, $\phi_{\rm 0}$ and $\psi_{\rm 0}$ satisfy
\begin{align}
& \tau_{\phi} \phi_{\rm 0} + \phi_{\rm 0}^3 - \mu_{\phi} - \Lambda \psi_0 = 0,
\nonumber \\
& \tau_{\psi} \psi_{\rm 0} + \psi_{\rm 0}^3 - \mu_{\psi} - \Lambda \phi_0 = 0.
\label{equi}
\end{align}
Expanding the free energy, Eq.~(\ref{freeenergy}),
and retaining the quadratic order terms
in $\delta \phi$ and $\delta \psi$, we obtain the Gaussian free-energy
\begin{align}
F_{\rm G}[\delta \phi, \delta \psi] = \int {\rm d}^2 r
\bigg[ 2 B  (\nabla^2 \delta \phi)^2 - 2 A (\nabla \delta \phi)^2
+\frac{{\epsilon}_{\phi}}{2} (\delta \phi)^2
\nonumber \\
+ 2 D (\nabla^2 \delta \psi)^2 - 2 C( \nabla \delta \psi)^2
+\frac{{\epsilon}_{\psi}}{2} (\delta \psi)^2
- \Lambda (\delta \phi) (\delta \psi) \bigg],
\label{free2}
\end{align}
where the notations
${\epsilon}_{\phi} = \tau_{\phi} + 3 \phi^2_{\rm 0}$ and
${\epsilon}_{\psi} = \tau_{\psi} + 3 \psi^2_{\rm 0}$ as well as
Eq.~(\ref{equi}) are used.

The static partial structure factor is
$S_{\phi \phi}(\mathbf{q}) =\langle \delta \phi(\mathbf{q})
\delta \phi(-\mathbf{q})\rangle$, and because of the radial symmetry,
$S_{\phi \phi}(\mathbf{q})=S_{\phi \phi}({q})$, where
$q=|\mathbf{q}|$.
It is given by
\begin{equation}
S_{\phi \phi}({q}) = \frac{2g_{\psi}(q)}
{4g_{\phi}(q) g_{\psi}(q) - \Lambda^2},
\label{shh}
\end{equation}
and similarly for $S_{\psi\psi}$ and $S_{\phi\psi}$
\begin{equation}
S_{\psi \psi}({q}) = \frac{2g_{\phi}(q)}
{4g_{\phi}(q) g_{\psi}(q) - \Lambda^2},
\label{sss}
\end{equation}
\begin{equation}
S_{\phi \psi}({q}) = S_{\psi \phi}({q}) =
\frac{\Lambda} {4g_{\phi}(q) g_{\psi}(q) - \Lambda^2},
\label{shs}
\end{equation}
and
\begin{align}
& g_{\phi}(q) = 2 B q^4 - 2 A q^2 +
\frac{{\epsilon}_{\phi}}{2},
\nonumber \\
& g_{\psi}(q) = 2 D q^4 - 2 C q^2 +
\frac{{\epsilon}_{\psi}}{2}.
\label{Ghs}
\end{align}
Since the structure factors diverge at $T_{\rm c}$, we see that the
coupling parameter $\Lambda$  effectively shifts the critical temperature to lower values.

\subsection{Decoupled leaflets ($\Lambda=0$)}

When the two leaflets are decoupled,
$\Lambda=0$,
it is sufficient to present results for only one of the two monolayers,
say the $\phi$ one.
From Eq.~(\ref{shh}), the decoupled structure factor is simply given by
\begin{equation}
S_{\phi\phi}({q}) = \frac{1}{2g_{\phi}(q)}.
\label{scat}
\end{equation}
This function has a peak at $q_{\phi}^{\ast}=\sqrt{A/2B}$ where
$g_{\phi}$ has a minimum, indicating that monolayer fluctuations have
a characteristic length scale describing their spatial modulations.
The peak height is hence given by
\begin{equation}
S_{\phi\phi}(q_{\phi}^{\ast}) = \frac{1}{{\epsilon}_{\phi} - A^2/B},
\label{heights}
\end{equation}
and diverges at ${\epsilon}_{\phi}^* = A^2/B$. In addition, the  structure factor in Eq.~(\ref{scat}) decays
as $\sim q^{-4}$ for large wavenumbers.

\begin{figure}[tbh]
\begin{center}
\includegraphics[scale=0.35]{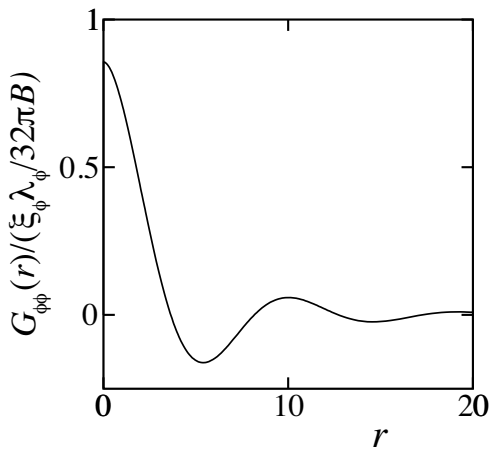}
\end{center}
\caption{
The rescaled (dimensionless) real-space correlation function, $G_{\phi\phi}$, of a single
monolayer in the structured-disordered phase as a function of $r$.
The parameters are $B/{\epsilon}_{\phi}=1$ and $\gamma_{\phi} = -0.9$, and the two resulting
characteristic lengths are $\lambda_\phi/2\pi\simeq 1.45$ and $\xi_\phi\simeq 6.32$.
}
\label{fig2}
\end{figure}

The correlation function
$G_{\phi\phi}(\mathbf{r})=\langle \delta \phi(\mathbf{r}) \delta
\phi(0)\rangle=G_{\phi\phi}({r})$, where $r= \vert \mathbf{r}\vert$,
is obtained by the inverse 2D Fourier transform of
Eq.~(\ref{scat}) (see Appendix A for the derivation)
\begin{equation}
G_{\phi\phi}({r}) = \frac{\xi_{\phi} \lambda_{\phi}}{32 \pi B}
{\rm Re} \left[ H_0^{(1)} \left( \frac{2 \pi r}{\lambda_{\phi}}
+ i \frac{r}{\xi_{\phi}} \right) \right],
\label{2d_corr}
\end{equation}
where $H_0^{(1)}(z)$ is the zeroth order Hankel function of the first kind,
and ``Re'' stands for the real part.
This correlation function contains two length scales; the first
is the modulation periodicity
\begin{equation}
\frac{\lambda_{\phi}}{2 \pi} =
\left(\frac{B}{{\epsilon}_{\phi}}\right)^{1/4} \frac{2}{\sqrt{\
1 - \gamma_{\phi}}},
\label{periodicity}
\end{equation}
and the second is the correlation length
\begin{equation}
\xi_{\phi} = \left(\frac{B}{{\epsilon}_{\phi}}\right)^{1/4}
\frac{2}{\sqrt{1 + \gamma_{\phi}}},
\label{corrleng}
\end{equation}
where
$\gamma_{\phi} = -A / \sqrt{{\epsilon}_{\phi} B}$.

In Fig.~\ref{fig2}, we plot Eq.~(\ref{2d_corr}) for certain parameter values.
The correlation function
has exponentially decaying oscillations as a function of the distance $r$,
similar to 3D microemulsions~\cite{GSbook}.
When $-1 \le \gamma_{\phi} \le 1 $, both $\lambda_{\phi}$ and $\xi_{\phi}$
are finite, and the corresponding phase is called the
\textit{structured-disordered} phase.
The peak of the structure factor $S_{\phi \phi}$ occurs at
$q^* > 0$ for $-1 \le \gamma_{\phi} \le 0$, whereas $q^* = 0$ for
$0 \le \gamma_{\phi} \le 1$.
The line at $\gamma_{\phi}=0$ is the \textit{Lifshitz line}.
The modulation periodicity $\lambda_{\phi}$ diverges for
$\gamma_{\phi} \to 1$, and the monolayer transforms into a
\textit{disordered} phase for $\gamma_{\phi} > 1$.
The line of $\gamma_{\phi}=1$ is called the \textit{disorder line}
and is not a phase transition line.
On the other hand, the correlation length $\xi_{\phi}$ diverges for
$\gamma_{\phi} \to -1$ (hence, this is the critical point), and the
\textit{ordered} phase appears for $\gamma_{\phi} < -1$~\cite{GSbook}.

As shown in Fig.~\ref{fig3}(a), the phase diagram of a decoupled bilayer
is easily obtained by combining the phase sequences in terms of
the two independent parameters
$\gamma_{\phi}=- A/\sqrt{{\epsilon}_{\phi}B}$
and $\gamma_{\psi}=- C/\sqrt{{\epsilon}_{\psi}D}$.
The various phases are expressed by the binary combination of ordered
(O), structured-disordered (S) and disordered (D) phases for each
of the leaflets.
The two disorder lines are shown in the figure by  dashed
lines, while the Lifshitz lines occur at $\gamma_\phi=0$ and $\gamma_\psi=0$
(not shown).

\subsection{Coupled leaflets ($\Lambda \neq 0$)}

\begin{figure}[tbh]
\begin{center}
\includegraphics[scale=0.35]{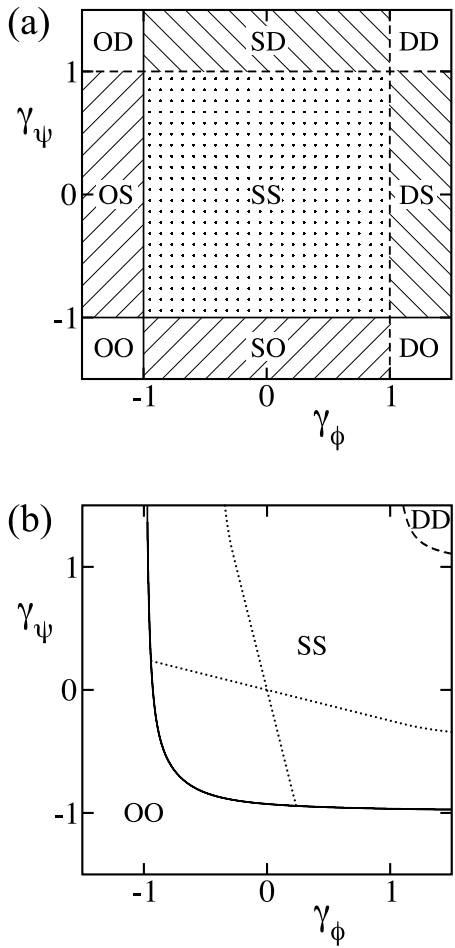}
\end{center}
\caption{
(a) Phase diagram of a decoupled modulated bilayer ($\Lambda = 0$)
plotted in the plane of
$\gamma_{\phi} = -A / \sqrt{{\epsilon}_{\phi} B}$ and
$\gamma_{\psi} = -C / \sqrt{{\epsilon}_{\psi} D}$.
The phases are labeled using a binary combination of two letters
representing phases in the $\phi$ and $\psi$ monolayers.
``O'', ``S'' and ``D'' stand for the ordered, structured-disordered
and disordered phases, respectively.
Solid and dashed lines are the phase transition and
disorder lines, respectively.
(b) Phase diagram of a coupled modulated bilayer ($\Lambda=1$)
for ${\epsilon}_{\phi} = {\epsilon}_{\psi} = 2$, $B=D=0.5$. In addition, the two dotted lines are the Lifshitz lines.
}
\label{fig3}
\end{figure}

When the two monolayers are coupled ($\Lambda \neq 0$), the
corresponding phase diagram can be obtained by analyzing the poles
in the complex plane of Eqs.~(\ref{shh})--(\ref{shs}).
One example is shown in Fig.~\ref{fig3}(b).
The boundary between the OO and SS phases (solid line) is determined
by the condition that all the structure factors diverge.
In the DD phase, all the poles are pure imaginary, whereas in the SS phase the poles
are complex, and at least one real pole exists in the OO phase.

\begin{figure}[tbh]
\begin{center}
\includegraphics[scale=0.35]{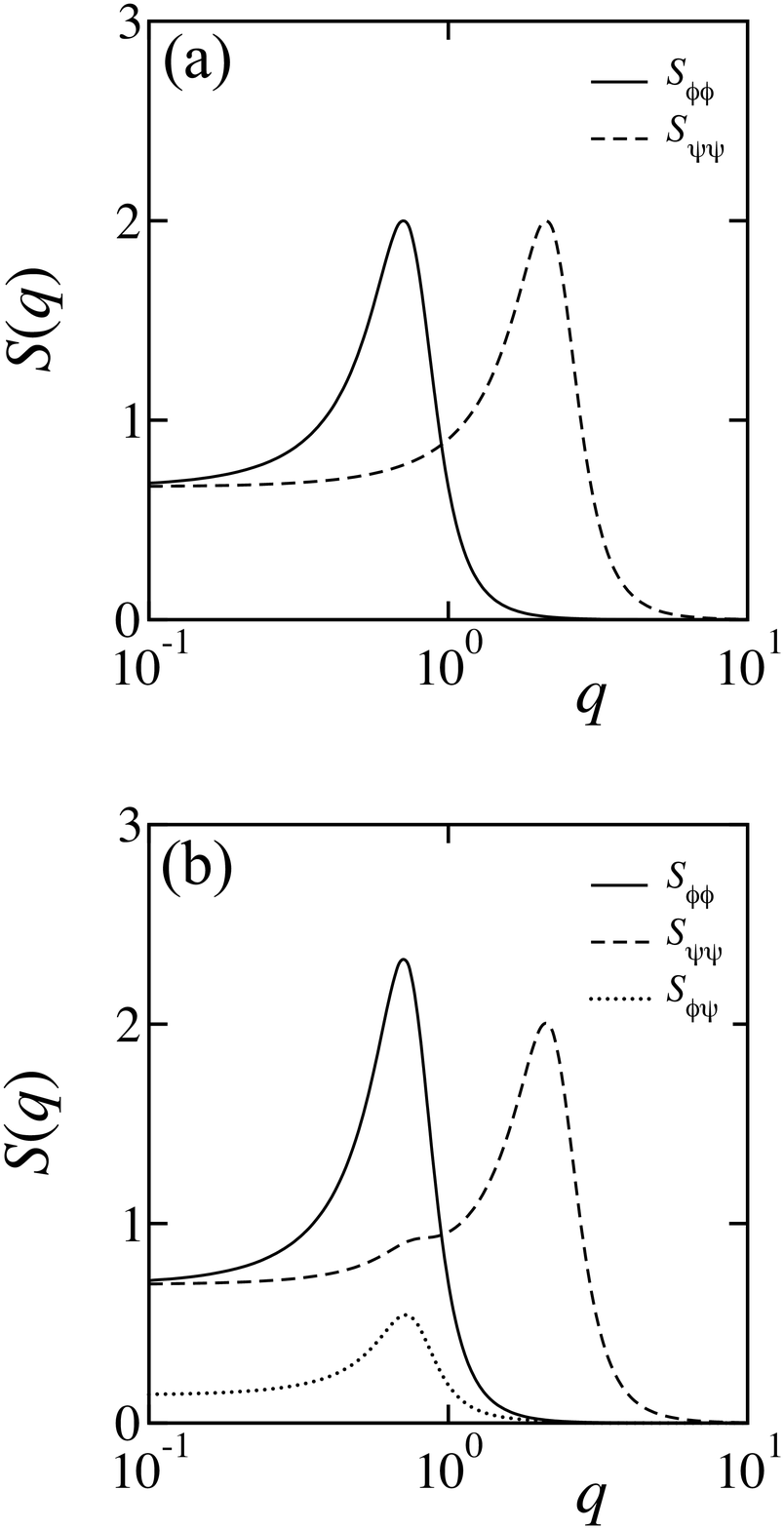}
\end{center}
\caption{
Bilayer structure factors $S_{\phi\phi}$, $S_{\psi\psi}$, and
$S_{\phi\psi}$ as a function of the wavenumber $q$ for
${\epsilon}_{\phi}={\epsilon}_{\psi}=1.5$,
$B=A=1$, $D=0.0123$, $C=0.1111$
($q^*_{\phi} = 1/\sqrt{2}, q^*_{\psi}=3/\sqrt{2}$).
(a) the decoupled case, $\Lambda = 0$; (b) the coupled case with
$\Lambda = 0.3$.
}
\label{fig4}
\end{figure}

In Fig.~\ref{fig3}(b) we see that the asymmetric phases such as DO and
SD are suppressed as compared with Fig.~\ref{fig3}(a).
This is because one of the leaflets induces modulations in the other
leaflet due to the inter-leaflet coupling, $\Lambda$.
In addition, the OO and SS phases are expanded as compared to
Fig.~\ref{fig3}(a).
As mentioned before, increasing the coupling $\Lambda$
effectively lowers the bilayer temperature.
The Lifshitz lines (dotted lines) are obtained by plotting
the main peak positions of Eqs.~(\ref{shh}) and (\ref{sss}).
Due to the coupling effect, they are tilted as compared to the $\Lambda=0$
case where the lines occur at $\gamma_{\phi}=0$ and $\gamma_{\psi}=0$.

In Fig.~\ref{fig4}(a) and (b), we plot the structure factors of the
decoupled and coupled bilayers,
respectively.
As an illustration of the coupling effect with $\Lambda=0.3$, we consider two different
characteristic wavelengths $q_{\phi}^* \ne q_{\psi}^*$.
The heights of the two peaks are set to be equal by requiring
${\epsilon}_{\phi}={\epsilon}_{\psi}$ and
$A^2/B = C^2/D =1$ (see Eq.~(\ref{heights})).
The peak height of
$S_{\phi\phi}$ at $q_{\phi}^*=1/\sqrt{2}$ is increased (as in Fig.~\ref{fig4}(b)) due to the coupling effect,
whereas that of $S_{\psi\psi}$ at $q_{\psi}^*=3/\sqrt{2}$ is almost
unchanged compared with the decoupled case.
We also plot $S_{\phi\psi}$ given by Eq.~(\ref{shs}) which
represents the cross correlation of fluctuations between the two
monolayers.
This quantity is proportional to the coupling constant $\Lambda$ and
its peak position is essentially determined by that of $S_{\phi\phi}$
at $q_{\phi}^*=1/\sqrt{2}$.

\begin{figure}[tbh]
\begin{center}
\includegraphics[scale=0.35]{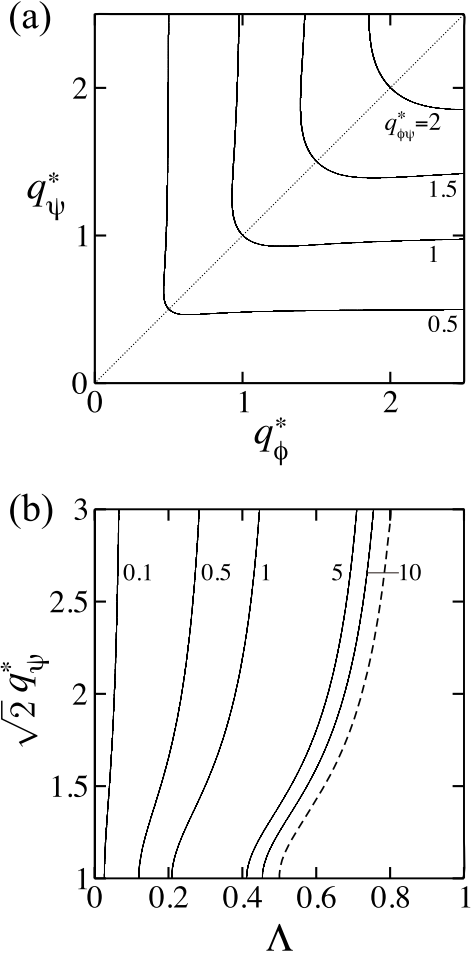}
\end{center}
\caption{
(a) Contour plot of the peak position $q_{\phi\psi}^*$ of the structure
factor $S_{\phi\psi}$ as function of $q_{\phi}^* = \sqrt{A/2B}$ and
$q_{\psi}^* = \sqrt{C/2D}$.
The parameters are $\Lambda = 0.3$,
${\epsilon}_{\phi}={\epsilon}_{\psi}=1.5$, and $A^2/B=C^2/D$.
The dotted line represents $q_{\phi}^*=q_{\psi}^*$.
(b) Contour plot of the peak height of the structure factor
$S_{\phi\psi}(q_{\phi\psi}^*)$ as a function of $\Lambda$ and
$\sqrt{2}q_{\psi}^*$. For the choice of parameters
$B=A=1$ and $C^2/D=1$, $q_{\phi}^*=1/\sqrt{2}$, the two characteristic wavenumbers satisfy
$q_{\psi}^*/q_{\phi}^*=\sqrt{2}q_{\psi}^*$.
The dashed line is the phase-transition line.
}
\label{fig5}
\end{figure}

The peak position of $S_{\phi\psi}$ denoted as $q_{\phi\psi}^*$
for the coupled case ($\Lambda=0.3$) is obtained numerically and
plotted in Fig.~\ref{fig5}(a) as a function of $q_{\phi}^*$ and
$q_{\psi}^*$, which are, respectively, the peak positions of $S_{\phi\phi}$ and
$S_{\psi\psi}$ for the decoupled case.
We find that for the coupled case, the value of $q_{\phi\psi}^*$
is almost equal to that of the smaller of $(q_{\phi}^*, q_{\psi}^*)$.
In Fig.~\ref{fig5}(b), the peak height of $S_{\phi\psi}$ at
$q_{\phi\psi}^*$ is plotted as a function of the coupling parameter
$\Lambda$ and the ratio of the two characteristic wavenumbers
$q_{\psi}^*/q_{\phi}^*$ for the specific case of $q_{\phi}^*=1/\sqrt{2}$,
as
the parameters are fixed to $B=A=1$ and
$C^2/D=1$.
The peak height increases as $\Lambda$ is increased because the
temperature of the coupled bilayer is effectively lowered.
To see this clearly, we have plotted in Fig.~\ref{fig5}(b) the phase-transition
line (dashed line) separating the structured-disordered and ordered phases.
The peak height of $S_{\phi\psi}$ also increases as
$q_{\psi}^*/q_{\phi}^*=\sqrt{2}q_{\psi}^*$ approaches unity, which is the case where
$S_{\phi \phi}$ and $S_{\psi \psi}$ completely overlap.

\subsection{Dynamic structure factors}

The dynamical fluctuations in composition, $\delta \phi(\mathbf{r},t)$
and $\delta \psi(\mathbf{r},t)$, depend on time $t$ and are now considered for
coupled modulated bilayers.
Since both $\delta \phi$ and $\delta \psi$ are conserved quantities
for each monolayer, the time evolution equations are given by
\begin{align}
& \frac{\partial \delta \phi (\mathbf{r}, t)}{\partial t} =
L_{\phi} \nabla^2 \frac{\delta F_{\rm G}}
{\delta (\delta \phi)} + \nu_{\phi} (\mathbf{r},t),
\nonumber \\
& \frac{\partial \delta \psi (\mathbf{r}, t)}{\partial t} =
L_{\psi} \nabla^2 \frac{\delta F_{\rm G}}
{\delta (\delta \psi)} + \nu_{\psi} (\mathbf{r},t),
\label{langevins}
\end{align}
where $L_{\phi}$ and $L_{\psi}$ are the kinetic coefficients taken to be
constants, and $\nu_{i} (\mathbf{r},t)$
represent Gaussian white noise, satisfying
$\langle\nu_{i} (\mathbf{r},t)\rangle = 0$ and
\begin{equation}
\langle\nu_{i} (\mathbf{r},t) \nu_{j} (\mathbf{r}',t')\rangle =
- \delta_{ij} L_{i} \nabla^2
\delta (\mathbf{r} - \mathbf{r}') \delta ( t - t'),
\end{equation}
where $i,j=\phi, \psi$ and $\langle \cdots \rangle$ now indicates the average over space and time.

The Fourier transform of fluctuations in both space and time is
\begin{equation}
\delta \phi (\mathbf{q},\omega) =
\int {\rm d}^2 r \, {\rm d}t \, \delta \phi (\mathbf{r},t)
e^{- i ({\mathbf{q}} \cdot \mathbf{r}-\omega t)},
\end{equation}
and a similar Fourier transform is used  for $\delta \psi$ and $\nu_{i}$.
The dynamic structure factors such as
$S_{\phi \phi}(\mathbf{q}, \omega) =\langle \delta \phi(\mathbf{q}, \omega)
\delta \phi(-\mathbf{q}, -\omega)\rangle$
are given by
\begin{equation}
S_{ij}({q}, \omega) =
\frac{\alpha_{ij}(q)}{\omega^2 + [\omega_+(q)]^2} +
\frac{\beta_{ij}(q)}{\omega^2 + [\omega_-(q)]^2},
\label{SF_q_o}
\end{equation}
where the explicit expressions of $\alpha_{ij}(q)$ and $\beta_{ij}(q)$ are
given in Table~\ref{table1} and the characteristic frequencies $\omega_{\pm}(q)$
are:
\begin{align}
& [\omega_{\pm}(q)]^2 =  \frac{1}{2}
\bigg[ \omega^2_{\phi}(q) + \omega^2_{\psi}(q)
\nonumber \\
& \mp \sqrt{[\omega^2_{\phi}(q) - \omega^2_{\psi}(q)]^2
+ 4 L_{\phi} L_{\psi} q^4 \Lambda^2 \omega^2_{\phi \psi}(q)} \bigg],
\label{omegapm}
\\
& \omega_{\phi}^2(q) = 4 L_{\phi}^2 q^4 [g_{\phi}(q)]^2+
L_{\phi} L_{\psi} q^4 \Lambda^2,
\label{omegaphi}
\\
& \omega_{\psi}^2(q) = 4 L_{\psi}^2 q^4 [g_{\psi}(q)]^2 +
L_{\phi} L_{\psi} q^4 \Lambda^2,
\label{omegapsi}
\\
& \omega_{\phi \psi}(q) = 2 q^2 [L_{\phi} g_{\phi}(q) +
L_{\psi} g_{\psi}(q)],
\label{omegaphipsi}
\end{align}
and $g_{\phi}$ and $g_{\psi}$ are defined in Eq.~(\ref{Ghs}).
In the more special case of $L_{\phi} = L_{\psi} = L $, $\omega_{\pm}$ reduce to a simpler form
\begin{align}
\omega_{\pm}(q) = L q^2 \biggl[& g_{\phi}(q)
+ g_{\psi}(q)
\nonumber \\
& \mp
\sqrt{(g_{\phi}(q) - g_{\psi}(q))^2 + \Lambda^2} \biggr].
\label{ome2}
\end{align}

\begin{table}[bht]
\caption{\label{table1}
Expressions for $\alpha_{ij}(q)$ and $\beta_{ij}(q)$}
\begin{ruledtabular}
\begin{tabular}{cc}
$\alpha_{ij}(q)$ & $\beta_{ij}(q)$ \\
\colrule
& \\
$\alpha_{\phi \phi}=\frac{\displaystyle 2 L_{\phi} q^2 (\omega^2_{\psi} - \omega^2_+)}{\displaystyle \omega_-^2 - \omega_+^2}$ &
$\beta_{\phi \phi}=\frac{\displaystyle 2 L_{\phi} q^2 (\omega^2_- - \omega^2_{\psi})}{\displaystyle \omega_-^2 - \omega_+^2}$ \\
& \\
$\alpha_{\psi \psi}=\frac{\displaystyle 2 L_{\psi} q^2 (\omega^2_{\phi} - \omega^2_+)}{\displaystyle \omega_-^2 - \omega_+^2}$ &
$\beta_{\psi \psi}=\frac{\displaystyle 2 L_{\psi} q^2 (\omega^2_- - \omega^2_{\phi})}{\displaystyle \omega_-^2 - \omega_+^2}$ \\
& \\
$\alpha_{\phi \psi}=\frac{\displaystyle 2 L_{\phi} L_{\psi} q^4 \Lambda \omega_{\phi \psi}}{\displaystyle \omega_-^2 - \omega_+^2}$ &
$\beta_{\phi \psi}=- \frac{\displaystyle 2 L_{\phi} L_{\psi} q^4 \Lambda \omega_{\phi \psi}}{\displaystyle \omega_-^2 - \omega_+^2}$\\
\end{tabular}
\end{ruledtabular}
\end{table}

\begin{figure}[tbh]
\begin{center}
\includegraphics[scale=0.35]{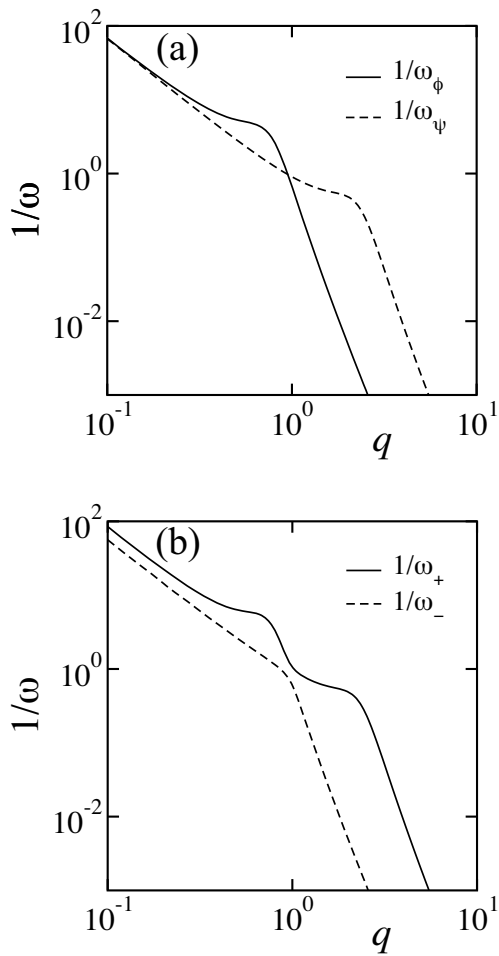}
\end{center}
\caption{
Decay time of concentration fluctuations as a function of the wavenumber
$q$ for ${\epsilon}_{\phi}={\epsilon}_{\psi}=1.5$,
$B=A=1$, $D=0.0123$, $C=0.1111$
(resulting in $q^*_{\phi} = 1/\sqrt{2}$ and $q^*_{\psi}=3/\sqrt{2}$), and $L = 1$.
(a) The decoupled case, $\Lambda=0$; (b) the coupled case with $\Lambda=0.3$.
}
\label{fig6}
\end{figure}

The intermediate structure factors $S_{ij}(q,t)$ depend
explicitly on time and are obtained by taking the inverse
Fourier transform in $t$:
\begin{equation}
S_{ij}({q}, t) =
\int \frac{\rm{d} \omega}{2 \pi} \, S_{ij}({q}, \omega)
e^{-i \omega t},
\label{def_Shh_qt}
\end{equation}
and from Eq.~(\ref{SF_q_o}), we obtain
\begin{equation}
S_{ij}({q}, t) = \frac{\alpha_{ij}(q)}{2 \omega_+(q)}
e^{- \omega_+(q) t}
+ \frac{\beta_{ij}(q)}{2 \omega_-(q)}
e^{- \omega_-(q) t}.
\label{Dynamic_stru}
\end{equation}
Hence the decay of concentration fluctuations is described by a
sum of two exponentials with two characteristic times
$1/\omega_{\pm}(q)$.
However, when the two monolayers are decoupled ($\Lambda=0$),
the $S_{\phi \phi}$ and $S_{\psi \psi}$ structure factors decay with
a single exponential characterized by a decay rate
$\omega_{\phi}$ and $\omega_{\psi}$ (see Eqs.~(\ref{omegaphi}) and
(\ref{omegapsi})), respectively.

For the decoupled case ($\Lambda=0$), the decay times $1/\omega_{\phi}$ and
$1/\omega_{\psi}$ are plotted as a function of $q$ in
Fig.~\ref{fig6}(a), with the same parameters as those in Fig.~\ref{fig4}.
The plots show a shoulder reflecting the characteristic structure at
wavenumbers $q_{\phi}^*=1/\sqrt{2}$ and $q_{\psi}^*=3/\sqrt{2}$.
Notice that the larger the initial length scale, the longer the decay time.
For the coupled case with $\Lambda=0.3$, we plot $1/\omega_{\pm}$
in Fig.~\ref{fig6}(b).
Due to the coupling effect, the two decay times split into a larger
and smaller one, $1/\omega_{+}>1/\omega_{-}$.
The larger one, $1/\omega_{+}$, exhibits two shoulders, while the
smaller one, $1/\omega_{-}$, has a shoulder between the two characteristic
wavenumbers.
The coupling affects the decay time of the structure corresponding
to the smaller wavenumber (larger length),
similar to the effect seen in Fig.~\ref{fig4} for the static structure factor.

\section{Coupled modulated phases with different periodicities
$q_{\phi}^* \neq q_{\psi}^*$}
\label{sec:separation}

\begin{figure*}[tbh]
\begin{center}
\includegraphics[scale=0.65]{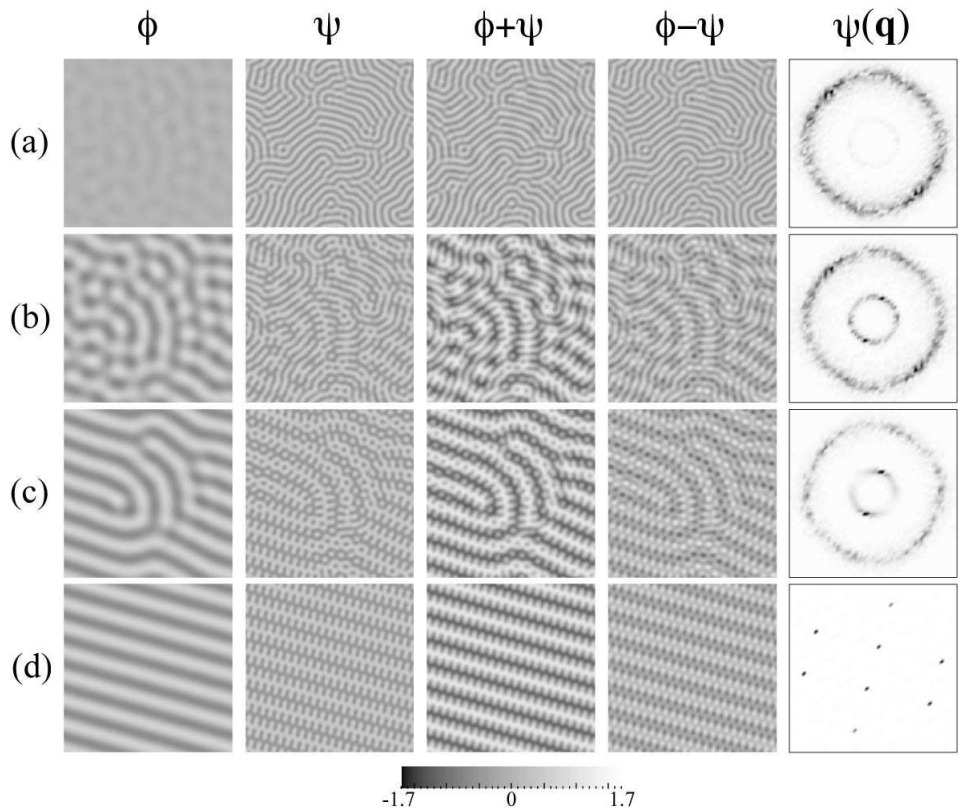}
\end{center}
\caption{
Time evolution of spatially modulated patterns in a coupled bilayer, which
consists of two stripe monolayers with different periodicities,
$q_{\phi}^* \neq q_{\psi}^*$.
Real-space patterns of $\phi$, $\psi$, $\phi+\psi$, $\phi-\psi$,
as well as the Fourier transform, $\psi({\mathbf q})$, are presented.
The time steps are: (a) $t=25$, (b) $t=60$, (c) $t=250$, (d) $t=5000$.
The parameters are $\phi_0 = \psi_0=0$, $\tau_{\phi}=\tau_{\psi}=0.8$,
$B=A=1$,  $D=0.0123$,  $C=0.1111$, $\Lambda=0.3$,
yielding $q_{\psi}^*/q_{\phi}^*=3$. In order to emphasize the color contrast,
the gray color code of the real-space patterns
is chosen to vary between $-1.7$ and $1.7$.
}
\label{figdynamics}
\end{figure*}

\begin{figure*}[tbh]
\begin{center}
\includegraphics[scale=0.65]{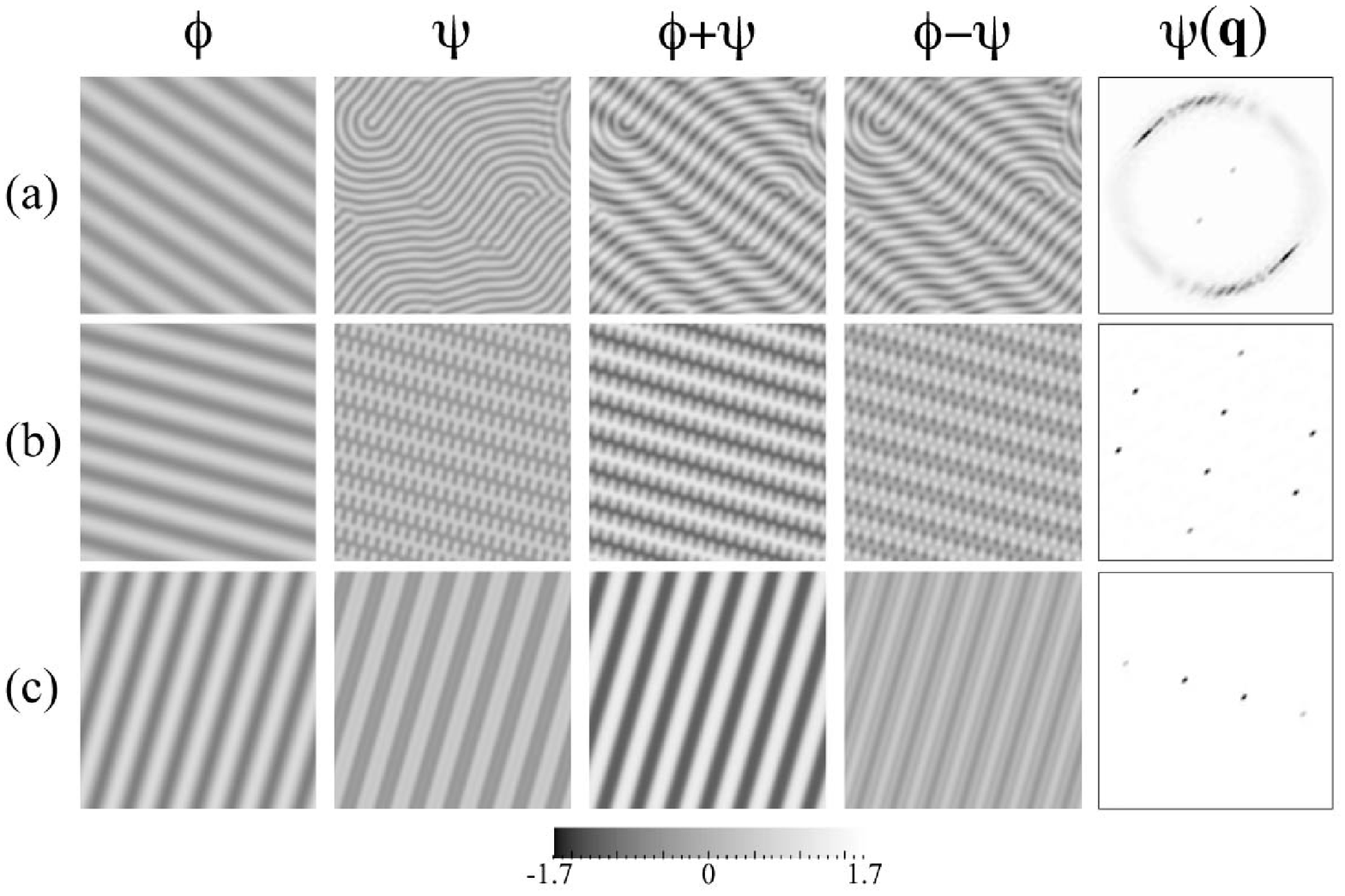}
\end{center}
\caption{
Equilibrated patterns (time step $t=5000$) of two coupled bilayers consisting of stripe
monolayers with different periodicities.
All parameters beside $\Lambda$ are the same as in Fig.~\ref{figdynamics}.
The $\Lambda$ parameter values are:
(a) $\Lambda=0.1$; (b) $\Lambda=0.3$; (c) $\Lambda=0.5$. In order to emphasize the color contrast, the color code
of the real-space patterns is chosen to vary between $-1.7$ and $1.7$.
}
\label{fig8}
\end{figure*}

In our previous paper~\cite{HKA}, we investigated the phase behavior
of coupled modulated bilayers by using the free energy,
Eq.~(\ref{freeenergy}), below $T_{\rm c}$.
When the two monolayers have the same preferred periodicity,
$q_{\phi}^* = q_{\psi}^*$, we obtained the mean-field phase diagram
exhibiting various combinations of modulated structures
such as stripe (S) and hexagonal (H) phases.
In some cases, the periodic structure in one of the monolayers induces
a similar modulation in the second monolayer.
 Moreover, the region of the induced modulated phase expands as the coupling parameter
$\Lambda$ becomes larger.

However, when the preferred periodicities in the two leaflets are different,
$q_{\phi}^* \neq q_{\psi}^*$, it is difficult to obtain the phase
diagram because the free energy densities cannot be obtained
analytically.
We then have to rely on numerical simulations \cite{HKA,Hirose11} to solve
the time evolution of the two coupled order-parameters as explained below.

For the dynamics of coupled modulated bilayers below $T_{\rm c}$,
we use the following time evolution equations~\cite{HKA}
\begin{align}
& \frac{\partial \phi(\mathbf{r}, t)}{\partial t} =
L_{\phi} \nabla^2 \frac{\delta F_{\rm b}}{\delta \phi},
\nonumber \\
& \frac{\partial \psi(\mathbf{r}, t)}{\partial t} =
L_{\psi} \nabla^2 \frac{\delta F_{\rm b}}{\delta \psi},
\label{conc_evol_s}
\end{align}
where the bilayer free-energy, $F_{\rm b}$, is given by
Eq.~(\ref{freeenergy}). Below $T_{\rm c}$ the effect of thermal fluctuations is less important
and the noise terms have been omitted in the above equations.
Hereafter, the kinetic coefficients $L_{\phi}$ and $L_{\psi}$ are
set to unity for simplicity.

We solve numerically the above 2D equations using periodic
boundary conditions.
Each run starts from a homogeneous state with a small random
noise around the average compositions $\phi_0$ and $\psi_0$.
Time is measured in discrete time steps, and $t=5000$ corresponds to
a well equilibrated system.
In all our simulations, the reduced temperature parameters are fixed to be
$\tau_{\phi}=\tau_{\psi}=0.8$.
The characteristic wavenumber in the $\phi$ monolayer is fixed as
$q_{\phi}^* = 1/\sqrt{2}$ by setting $B=A=1$, while the periodicity
in the $\psi$ monolayer $q^*_{\psi}=\sqrt{C/2D}$ is varied and
the condition $C^2/D=1$ is used as before.
In the following, we shall consider only the two coupled stripe phases
for $\phi_0=\psi_0=0$.
However, the combination of stripe and hexagonal phases leads to a
rich variety of complex patterns~\cite{Hirose11}.

\begin{figure}[tbh]
\begin{center}
\includegraphics[scale=0.4]{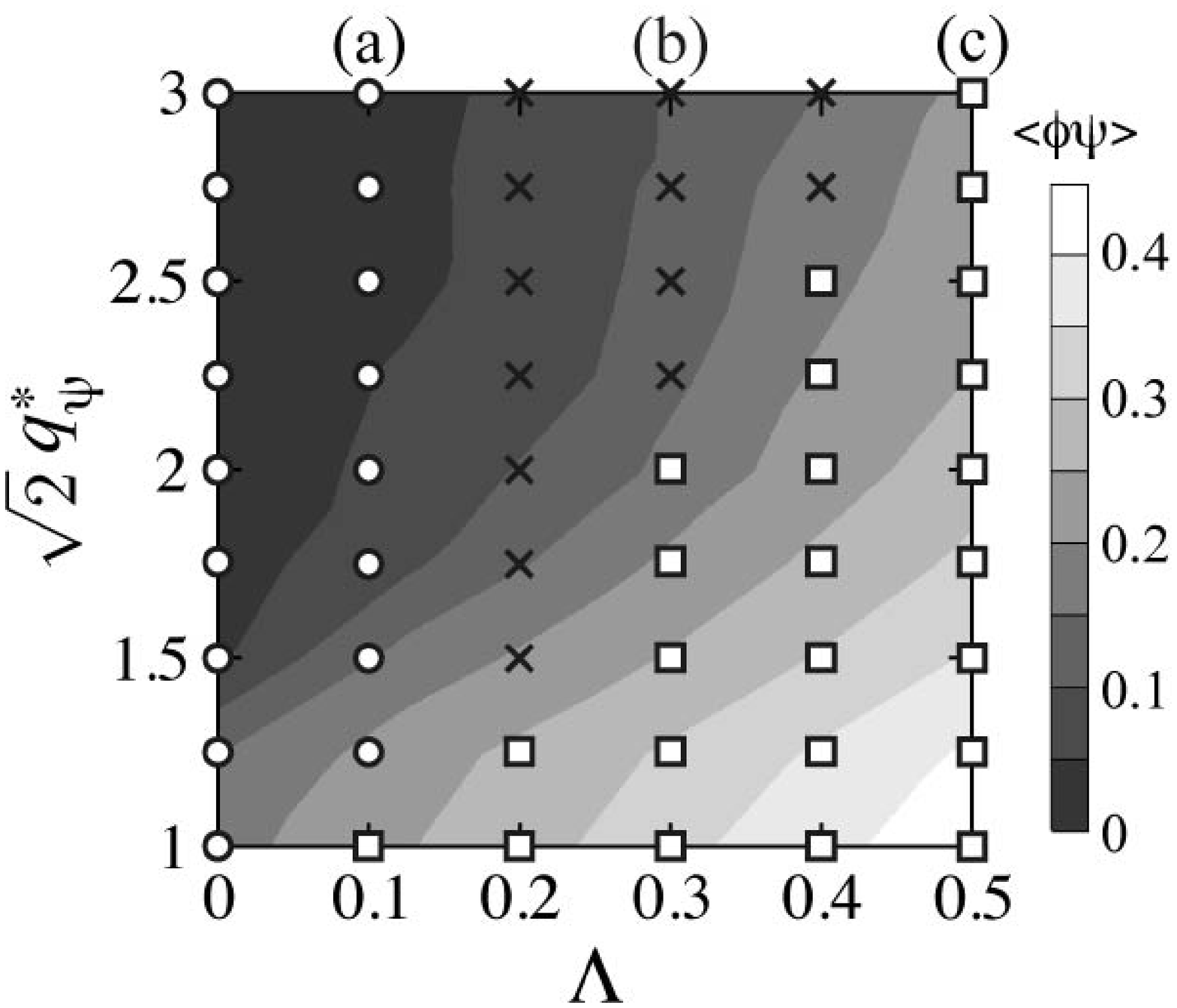}
\end{center}
\caption{
Contour plot of $\langle \phi \psi \rangle$ at $t=1000$
as a function of $\Lambda$ and $\sqrt{2}q_{\psi}^*$, for the specific case
of $q_\phi^*=1/\sqrt{2}$.
(a), (b), (c) correspond to $\Lambda=0.1, 0.3$ and $0.5$, respectively,
as given in Fig.~\ref{fig8}.
Circles, crosses, squares correspond, respectively, to independent,
intermediate, and coincident structures (see text).
}
\label{fig9}
\end{figure}

\begin{figure}[tbh]
\begin{center}
\includegraphics[scale=0.35]{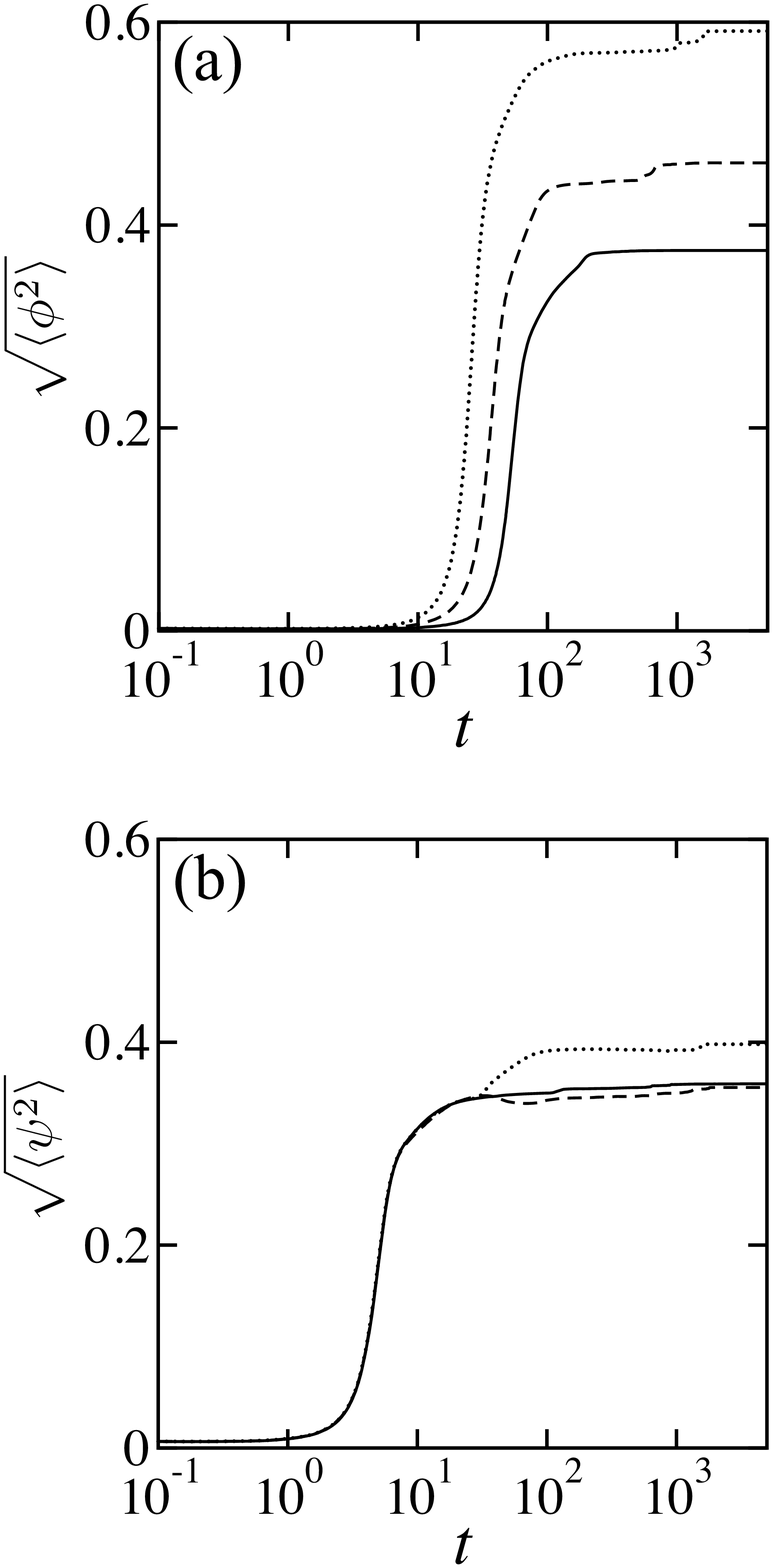}
\end{center}
\caption{
Time evolution of the modulation amplitudes of the two monolayers:
(a) $\sqrt{\langle \phi^2 \rangle}$ and
(b) $\sqrt{\langle \psi^2 \rangle}$.
The parameters are $\tau_{\phi}=\tau_{\psi}=0.8$, $\phi_0=\psi_0=0$,
$B=A=1$, $D=0.0123$, $C=0.1111$ yielding $q_{\psi}^*/q_{\phi}^*=3$.
The different lines represent $\Lambda=0.1$ (solid), $\Lambda=0.3$ (dashed), and
$\Lambda=0.5$ (dotted).
}
\label{figtime}
\end{figure}

When the characteristic wavenumbers of the two decoupled monolayers
are different, $q_{\phi}^* \neq q_{\psi}^*$, the two coupled
modulated structures cannot apparently match each other.
The frustration between the two different periodicities is
due to the inter-leaflet coupling and affects their morphologies.
In Fig.~\ref{figdynamics}, we show one example of the time evolution
of a coupled micro-phase separation for
$q_{\psi}^*/q_{\phi}^* = 3$ ($q_{\phi}^* = 1/\sqrt{2}$ and
$q_{\psi}^* = 3/\sqrt{2}$) and $\Lambda =0.3$.
The spatial patterns of $\phi$, $\psi$, $\phi+\psi$, $\phi-\psi$
and the Fourier transformed pattern, $\psi({\mathbf q})$, are
presented for time steps of $t=25, 60, 250$ and $5000$.
Starting from the isotropic state, the $\psi$ monolayer forms first
stripes (a).
Then, as the $\phi$ monolayer starts to segregate at larger $t$, it simultaneously chops
the $\psi$ stripes into smaller sections (b), and the $\psi$ monolayer
transforms into a finger-like patterns (c).
Reconnection of the $\phi$ stripes takes place after a long time of
annealing, and a pattern of alternating fingers in the $\psi$ monolayer
is finally obtained (d).
From the time evolution of the 2D Fourier patterns of the
$\psi$ monolayer, it is apparent that the intermediate structures
are characterized by the two length scales of ratio $1:3$.

In Fig.~\ref{fig8}, we show the spatially modulated patterns
at $t=5000$ of the two coupled $\phi$  and $\psi$ monolayers with different
periodicities ($q_{\phi}^* = 1/\sqrt{2}$ and $q_{\psi}^* = 3/\sqrt{2}$
as before) for different
values of the coupling parameter $\Lambda=0.1, 0.3$ and $0.5$.
Notice that Fig.~\ref{fig8}(b) is the same as
Fig.~\ref{figdynamics}(d).
In the weak-coupling case ($\Lambda=0.1$), the two monolayers
exhibit two independent stripe morphologies with characteristic
periodicities (called the ``independent'' morphology).
Here the two stripes essentially do not affect each other.
As the coupling constant is increased ($\Lambda=0.3$), stripes
with a finger-like structure appear in the $\psi$ monolayer, while
the stripe morphology in the $\phi$ monolayer is almost unaffected
(called the ``intermediate'' morphology).
In the Fourier transformed pattern of $\psi$, we clearly see that
the structures with two different characteristic wavenumbers are
coexisting.
This result is also in accord with the properties of the static
structure factors, $S_{ij}(q)$, having two different characteristic lengths,
as shown in Fig.~\ref{fig4}.
For a larger coupling parameter ($\Lambda=0.5$), very similar
patterns are obtained for $\phi$ and $\psi$, and almost coincide
with one another (called the ``coincident'' morphology).
It should be noted that the structure with the larger wavelength
dominates when the coupling is large enough, and
the sequence of morphological
changes shown in Fig.~\ref{fig8}, as $\Lambda$ increases, is rather typical.

In order to quantify the three morphologies (independent, intermediate,
coincident), we calculate the spatial average of the product of the two
compositions
\begin{equation}
\langle \phi \psi \rangle = \frac{1}{\cal{A}} \int {\rm d}^2 r \,
\phi(\mathbf{r},t)\psi(\mathbf{r},t),
\label{phipsicorr}
\end{equation}
where $\cal{A}$ is the total system area.
In Fig.~\ref{fig9}, we plot $\langle \phi \psi \rangle$ at $t=1000$
(sufficient for equilibration) for various combinations of $\Lambda$ and
$q_{\psi}^* /q_{\phi}^* = \sqrt{2} q_\psi^*$ (for the case $q_\phi^*=1/\sqrt{2}$).
The morphology of the obtained patterns is marked on the figure by circles, crosses and squares.
The values of $\langle \phi \psi \rangle$ are small for ``independent''
(circles) structures, while they become larger for the ``intermediate''
(crosses) and ``coincident'' (squares) morphologies.
Although the morphological changes are gradual and do not represent a sharp
transition, the intermediate patterns appear roughly for
$0.05 < \langle \phi \psi \rangle < 0.2$.
The region of intermediate structure expands as $q_{\psi}^* /q_{\phi}^*=\sqrt{2} q_\psi^*$
increases,  and the patterns coincide for $\Lambda=0.5$.
We note that although the morphology cannot be solely determined by
the quantity $\langle \phi \psi \rangle$, the behavior of $\langle \phi \psi \rangle$ is similar
to that of the peak values of the cross correlation,
$S_{\phi\psi}(q_{\phi\psi}^*)$, presented in Fig.~\ref{fig5}(b).

For the quantitative argument concerning the micro-phase separation
dynamics in each monolayer, we have also calculated the two self
quantities:
\begin{equation}
 \langle \phi^2 \rangle = \frac{1}{\cal{A}} \int {\rm d}^2r \,
\phi^2(\mathbf{r},t),
\end{equation}
and
\begin{equation}
 \langle \psi^2 \rangle = \frac{1}{\cal{A}} \int {\rm d}^2r \,
\psi^2(\mathbf{r},t).
\end{equation}
In Fig.~\ref{figtime}, we plot the square root of these quantities as
a function of $t$, with the same parameters as those used in
Fig.~\ref{fig8}.
In all studied cases, the modulation
of the $\psi$ monolayer having a larger wavenumber ($q_{\psi}^*>q_{\phi}^*$),
grows faster than that of
the $\phi$ monolayer.
We also remark that the structure formation in the $\phi$ monolayer
is accelerated for larger $\Lambda$, whereas that of the $\psi$ monolayer
is almost unchanged.
According to the linear stability analysis of Eq.~(\ref{conc_evol_s}),
the initial growth rates of the unstable modes are essentially equivalent
to the decay rates $\omega_{\pm}$ of the concentration fluctuations given
in Eq.~(\ref{ome2}).
This is consistent with Fig.~\ref{fig6}, where the characteristic growth
time ($1/\omega$) for
larger $q$ is smaller than that for smaller $q$.
The growth rates increases with $\Lambda$ (see Fig.~\ref{figtime}(a))
because the coupling effectively reduces the temperature and enhances
the phase transition.
The faster the decay of concentration fluctuations, the faster the
structure formation.

\section{Discussion and final remarks}
\label{sec:discussion}

In this paper we present a model for coupled modulated lipid
bilayers.
We start by considering a monolayer consisting of a mixture
of saturated and hybrid lipids, and
propose a phenomenological model that includes a coupling
between the lipid composition and a 2D vectorial field.
This coupling arises from the line-active nature of the hybrid lipid,
which adjusts its tail orientation in order to reduce the line tension.
Minimization of the monolayer free-energy with respect to the vectorial
field yields a 2D microemulsion exhibiting modulated phases.
The characteristic wavelength of modulation is determined by the monolayer
coefficients $A$ and $B$, Eq.~(\ref{e8}), reflecting molecular properties of lipid mixture.
We then construct a model for lipid bilayers comprised of two
modulated monolayers that influence each other through an inter-leaflet
coupling.

Based on the model, we study concentration fluctuations
of bilayers above $T_{\rm c}$, and calculate their
static structure factors.
The calculated phase diagram for coupled bilayers shows that
the extent of ordered and structured-disordered phases become larger
as compared to the decoupled case.
When the two monolayers with different preferred wavenumbers
$q_{\phi}^* \neq q_{\psi}^*$ are coupled, (say $q_{\phi}^* < q_{\psi}^*$), 
the peak height of $S_{\phi\phi}$ occurring at smaller
$q$-numbers becomes larger as compared to the decoupled case, 
whereas the peak height of $S_{\psi\psi}$
occurring at larger $q$-numbers almost
does not change.
Namely, the inter-leaflet coupling strongly affects the
compositional modulation in each monolayer.
Furthermore, the inter-leaflet coupling has a clear signature on
the cross structure factor, $S_{\phi \psi}$, as well as
on the dynamics of
concentration fluctuations.
By calculating the intermediate structure factor, $S(q,t)$, we show
that concentration fluctuations generally exhibit a double exponential
decay with two decay rates, $\omega_\pm$.
One of the decay times ($1/\omega_+$) exhibits two shoulders at 
wavenumbers describing the monolayer compositional modulations.

For membranes below $T_{\rm c}$, we studied the micro-phase separation
of a coupled modulated bilayer.
When the two monolayers have different modulations,
$q_{\phi}^* \neq q_{\psi}^*$, we obtained numerically
a variety of complex patterns.
The initial growth rates of the unstable modes are identical
to the decay rates of the concentration fluctuations.

As mentioned in Sec.~I, the special character of concentration fluctuations
in our model may
explain the finite-size domains (``rafts") in biological membranes.
For ordinary binary mixtures above $T_{\rm c}$, the only length scale of the disorder phase
is determined by the correlation length, and
close to $T_{\rm c}$, this length becomes large.
But within our 2D microemulsion model, there is another length scale
characterizing the modulations as given by Eq.~(\ref{periodicity}).
This second length scale may also explain the finite-size domains observed
in some experiments as a result of micro-phase separation in the low
temperature phase.

A related model based on a microemulsion picture was recently
proposed by Schick~\cite{Schick}, who considered a coupling between curvature and
compositional asymmetry between the two leaflets, resulting in a 2D microemulsion.
Although Schick's model as well as ours share the microemulsion viewpoint,
the origin of the physical mechanism is different.
In Schick's model the coupling between composition asymmetry and curvature gives rise
to domains with different spontaneous curvature.
Our model, on the other hand,  assumes that for flat monolayers, the line-active nature of hybrid lipids
is solely responsible for the microemulsion formation (see Eq.~(\ref{freeenergycoup})).
Hence, our proposed physical mechanism for
the bilayer coupling, as discussed in Sec.~\ref{couplebilayer}, is different.
One of the consequences of our $\Lambda>0$ coupling is
that the domains residing on the two leaflets are in register,
unlike the prediction of Ref.~\cite{Schick}.

The present work is concerned with the analysis of equilibrium
properties and relaxation dynamics towards the
equilibrium state.
The existence of finite-size domains may also be explained by
non-equilibrium lipid transport between the cell interior and the
membrane.
Such a mechanism was considered~\cite{Foret,GSR08,GSR09,FSH2}
through a coupling between the
membrane and an outer reservoir of lipid or cholesterol.
Similarly to our model, these works explain the
appearance of the finite-size domains as a result of micro-phase
separation.

Asymmetry between lipid composition in the inner and outer leaflets
of biomembranes has a very
deep significance and is closely
related to the cell biological functions.
For instance, the breakdown of such compositional asymmetry is
related to programmed cell death (apoptosis)~\cite{MWS}.
For living cells, this asymmetry
is maintained by an enzyme called ``flippase'' that actively
transports the lipids between the two leaflets~\cite{Daleke}.
The half-time of lipid composition due to flip-flop motion is
measured using  time resolved small angle neutron
scattering (SANS)~\cite{NFKEH}, and is estimated to be several hours
at physiological conditions.

Although membranal signal transduction is important for
various biological functions, its dynamical properties are still
not so well understood.
As our theory offers an explanation for the size and dynamics of lipid
domains, we hope that it and similar models will contribute in the future towards
the understanding of functional
processes in biomembranes.

\begin{acknowledgments}

We thank T.\ Kato, S.\ L.\ Keller, S.\ Ramachandran, M.\ Schick and
K.\ Yamada for useful discussions.
YH acknowledges a Research Fellowship for Young Scientists
No.\ 215097 from the JSPS.
This work was supported by Grant-in-Aid for Scientific Research
(grant No.\ 21540420) from the MEXT of Japan.
SK acknowledges support by the JSPS Core-to-Core Program
{\it ``International research network for non-equilibrium dynamics
of soft matter''}.
DA acknowledges support from the Israel Science Foundation (ISF)
under grant No.\ 231/08 and the US-Israel Binational Foundation (BSF)
under grant No.\ 2006/055.
\end{acknowledgments}

\appendix
\section{Derivation of $G_{\phi\phi}(r)$}

In this Appendix we present the derivation of $G_{\phi\phi}$ of Eq.~(\ref{2d_corr}).
The real-space correlation function is given by the 2D inverse Fourier
transform of Eq.~(\ref{scat}):
\begin{align}
G_{\phi\phi}({r}) &= \int \frac{{\rm d}^2 q}{(2\pi)^2} \,
S_{\phi\phi}({q}) e^{i \mathbf{q} \cdot \mathbf{r}}
\nonumber \\
     &= \frac{1}{4\pi B} \int^{\infty}_0 {\rm d} q \,
\frac{q J_0(q r)}{q^4-(A/B)q^2+ {\epsilon}_{\phi}/4B},
\label{2dcorr}
\end{align}
where $J_0(qr)$ is the zeroth-order Bessel function.
We use the relation
$J_0(qr) = [H_0^{(1)}(qr)+H_0^{(2)}(qr)]/2$
where $H_0^{(1)}(qr)$ and $H_0^{(2)}(qr)$ are the zeroth-order Hankel
functions of the first and second kind, respectively.
Then, the integral in Eq.~(\ref{2dcorr}) is written as
$I=(I_1+I_2)/2$ where
\begin{equation}
I_{i} = \int^{\infty}_0 {\rm d} q \,
\frac{q H_0^{(i)}(q r)}{q^4-(A/B)q^2+ {\epsilon}_{\phi}/4B},
\end{equation}
with $i=1,2$.

\begin{figure}[bht]
\begin{center}
\includegraphics[scale=0.35]{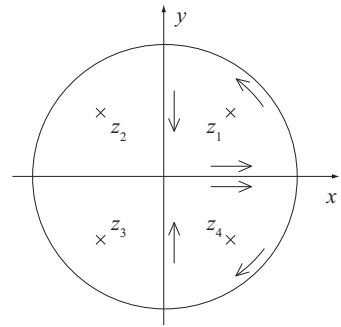}
\end{center}
\caption{
Four poles in the complex plane at $z=z_j$ ($j=1,2,3$ and $4$).
The two close integral paths are indicated by the arrows in the 1st and 4th quadrants.
}
\label{complexfig}
\end{figure}

In order to evaluate the above integral, we performed the integration in the
complex plane by replacing $q$ with the complex variable $z=x+iy$.
The integrand has four poles at
\begin{eqnarray}
z_j &=& \frac{({\epsilon}_{\phi}/B)^{1/4}}{2}
\left( \pm \sqrt{1-\gamma_{\phi}} \pm i \sqrt{1+\gamma_{\phi}}\right) \nonumber\\
& =& \pm\frac{2\pi}{\lambda_\phi}~\pm i\frac{1}{\xi_\phi} \, ,
\label{4poles}
\end{eqnarray}
with $j=1, 2, 3, 4$ and $\lambda_\phi$ and $\xi_\phi$ have been defined in
Eqs.~(\ref{periodicity}) and (\ref{corrleng}).
These poles are located in the quadrants 1, 2, 3 and 4, off the real
and imaginary axes.
For the integral $I_1$, we integrate along a quarter-circle contour in the first quadrant
in an anti-clockwise direction. The contour radius is taken
to infinity (Fig.~\ref{complexfig}), and
using the residue theorem we obtain
\begin{align}
&\int_0^{\infty} {\rm d}x \,
\frac{x H_0^{(1)}(xr)}{x^4-(A/B)x^2+{\epsilon}_{\phi}/4B}
\nonumber \\
&+ \int_{\infty}^0 {\rm d}y \,
\frac{iy H_0^{(1)}(iyr)}{y^4+(A/B)y^2+{\epsilon}_{\phi}/4B}
= 2 \pi i \, {\rm Res}\vert_{z_1},
\label{int1}
\end{align}
where ``Res'' stands for the residue.
Similarly, for the integral $I_2$, we integrate along the contour of the
quarter-circle of infinite radius in the fourth quadrant in the
clockwise direction:
\begin{align}
& \int_0^{\infty} {\rm d}x \,
\frac{x H_0^{(2)}(xr)}{x^4-(A/B)x^2+{\epsilon}_{\phi}/4B}
\nonumber \\
& + \int_{-\infty}^0 {\rm d}y \,
\frac{iy H_0^{(2)}(iyr)}{y^4+(A/B)y^2+{\epsilon}_{\phi}/4B}
= - 2 \pi i \,{\rm Res}\vert_{z_4}.
\label{int4}
\end{align}

Combining Eqs.~(\ref{int1}) and (\ref{int4}), and further using the
relation $H_0^{(1)}(-z)=-H_0^{(2)}(z)$, we obtain
\begin{equation}
I = \frac{\pi}
{2\sqrt{{\epsilon}_{\phi}/B} \sqrt{1 - \gamma_{\phi}^2}}
\left[ H_0^{(1)} (z_1r)+ H_0^{(2)}(z_4r) \right].
\end{equation}
Finally, using $\lambda_{\phi}$ and $\xi_{\phi}$ from Eq.~(\ref{4poles}),
we obtain
\begin{equation}
G_{\phi\phi}({r}) =
\frac{\xi_{\phi} \lambda_{\phi}}{32 \pi B}
{\rm Re} \left[ H_0^{(1)}
\left( \frac{2 \pi r}{\lambda_{\phi}} +
i \frac{r}{\xi_{\phi}} \right) \right].
\end{equation}
In the above expressions, we have used the relation
$H_0^{(2)}(\overline{z}) = \overline{H_0^{(1)}}(z)$, where
$\overline{z}$ is the complex conjugate of $z$.


\begin{thebibliography}{99}
\bibitem{footnote1}
In our model we always refer to the critical point, $T_{\rm c}$, at which the correlation length diverges. This is
related to the miscibility transition temperature
measured in experiments.
\end{thebibliography}

\end{document}